\documentclass[12pt,a4]{article}
\usepackage{epsfig}
\usepackage{amsfonts}
\title {\sc  One-Time Pad, Arithmetic Coding and Logic Gates: An unifying theme using Dynamical Systems}
\author{Nithin Nagaraj and Prabhakar G. Vaidya \\ School of Natural  Sciences and Engineering \\ National Institute of Advanced Studies \\ IISc Campus, Bangalore 560012 \\{\bf \small Email:~nithin\_nagaraj@yahoo.com}}
\date{February 22, 2008}

\begin{document}
\maketitle

{\center \section*{Abstract}}

\noindent In this letter, we prove that the perfectly secure
One-Time Pad (OTP) encryption can be seen as finding the initial
condition on the binary map under a random switch based on the
perfectly random pad. This turns out to be a special case of
Grangetto's randomized arithmetic coding performed on the Binary
Map. Furthermore, we derive the set of possible perfect secrecy
systems using such an approach. Since OTP encryption is an XOR
operation, we thus have a dynamical systems implementation of the
XOR gate. We show similar implementations for other gates such as
NOR, NAND, OR, XNOR, AND and NOT. The dynamical systems framework
unifies the three areas to which Shannon made foundational
contributions: lossless compression (Source Coding), perfect
encryption (Cryptography), and design of logic gates
(Computation).\\

\section{Shannon's Legacy: Coding, Cryptography and Computation}
Claude Shannon was one of the most important figures in the
information revolution of the last century. He made foundational
contributions to Coding, Cryptography and Computation. His master's
thesis used Boolean algebra to analyze and synthesize switching and
computer circuits~\cite{ShannonMS}. In 1948, he formulated a
mathematical theory of communication where among other things, he
was the first to define Entropy as the fundamental limit of
noiseless lossless source coding~\cite{Shannon48}. In the following
year, he proved the perfect secrecy of the Vernam cryptographic
system, also popularly known as One-Time Pad (OTP)~\cite{Shannon49}.
OTP is the only method to boast of perfect secrecy.

In this paper, we attempt to unify these three themes by a dynamical
systems framework. We claim that the three things are deeply related
when viewed from a dynamical systems perspective. The letter is
organized as follows. In Section 2 we introduce the Binary Map and
its skewed cousins, a piece-wise linear dynamical system which is
Lebesgue measure (in this case, this is the probability measure)
preserving, chaotic and ergodic. Section 3 introduces arithmetic
coding as finding the initial condition on the skewed Binary Map.
Section 4 introduces Grangetto's randomized arithmetic coding and
establishes its connection with the OTP. Section 5 deals with
generalizing OTP to higher alphabets. Section 6 talks about
implementation of logic gates using randomized arithmetic coding. We
conclude in Section 7.

\section{The Binary Map and its Skewed Cousins}
\begin{figure}[!h]
\centering
\includegraphics[scale=.6]{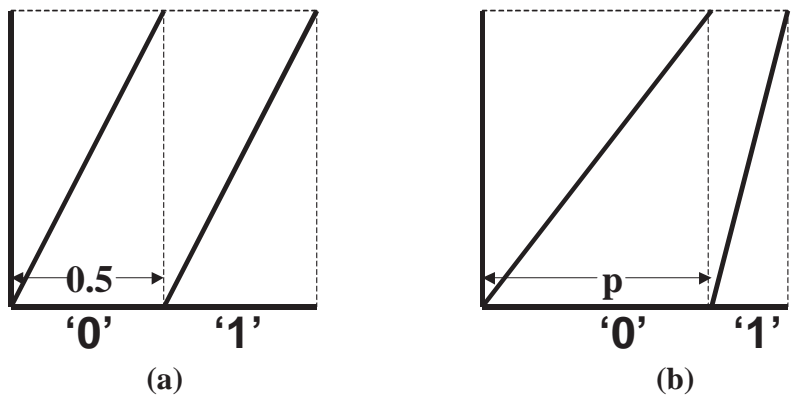}
\caption{(a) Binary Map. (b) Skewed Binary Map. Both these are
examples of GLS.} \label{fig:figbin1} \vspace{0.2in}
\end{figure}

The Binary Map (Fig.~\ref{fig:figbin1}(a)) $T:[0,1) \rightarrow
[0,1)$ is defined as:
\begin{eqnarray*}
x & \mapsto & 2x,~~~~~~~~~~~~~~~~~ 0 \leq x<\frac{1}{2}\\
    & \mapsto & 2x-1,~~~~~~~~~~~~ \frac{1}{2} \leq x < 1.
\end{eqnarray*}

It is well known that the binary map is a non-linear chaotic
dynamical system, which preserves the Lebesgue
measure~\cite{Dajani02}. Furthermore, every initial condition in
$[0,1)$ has a unique symbolic sequence and every finite length
($>0$) symbolic sequence corresponds to a subset of $[0,1)$ of
non-zero measure. Since the binary map has the maximum topological
entropy for two symbols ($=ln(2)$), all possible arrangements of 0
and 1 can occur in its space of symbolic sequences.

The symbol `0' corresponds to the interval $[0,0.5)$ and the symbol
`1' corresponds to the interval $[0.5,1)$. The binary map belongs to
a larger class of dynamical systems known as Generalized Lur\"{o}th
Series (GLS)~\cite{Dajani02}. GLS is studied for its number
theoretical properties. The skewed binary map is shown in
Fig.~\ref{fig:figbin1}(b). Here the symbols `0' and `1' correspond
to the intervals $[0,p)$ and $[p,1)$ respectively ($0 \leq p \leq
1$, $p=0.5$ corresponds to the binary map).

\subsection{Modes of Skewed Binary Map}
\begin{figure}[!h]
\centering
\includegraphics[scale=.6]{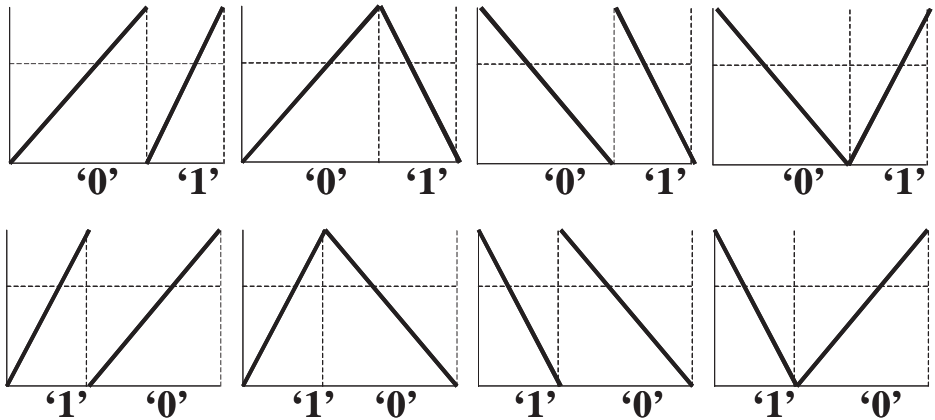}
\caption{Different modes of the skewed binary map. The bottom row
are duals of the maps in the first row.} \label{fig:figmodes}
\vspace{0.2in}
\end{figure}

There are 8 different modes of the skewed binary map as shown in
Fig.~\ref{fig:figmodes}. These are obtained by a combination of
swapping the two intervals corresponding to `0' and `1', and
changing the sign of the slope of the map in the two intervals. We
shall call a map with two alphabets a {\it dual} of the another if
the two intervals along with their symbols are swapped.

The GLS can be readily extended to larger alphabets.

\section{Arithmetic coding seen as a Dynamical System}
Recently, we have proposed a method of lossless data compression
where the (binary) message ($M$) is treated as the symbolic sequence
on the appropriate GLS ($p$ corresponds to the probability of the
alphabet `0' in the message) and the initial condition is determined
by iterating backwards. The initial condition serves as the
compressed file which can be used to determine the symbolic sequence
(message) at the decoder (given $p$). We called such a method as
GLS-coding. GLS-coding is a generalization of arithmetic coding
which achieves the Shannon's entropy rate for the source. Thus
GLS-coding yields optimal noiseless lossless data compression. For
full details please refer to \cite{Nagaraj08}.

\section{Randomized Arithmetic Coding and the One Time Pad}
\begin{figure}[!h]
\centering
\includegraphics[scale=.6]{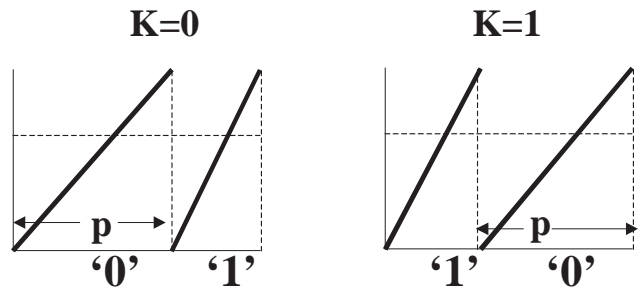}
\caption{Grangetto's Randomized Arithmetic Coding. `$K$' is the
binary key stream. When $K=0$, the backward iteration is done on the
skewed binary map on the left and for $K=1$, the backward iteration
is done on its dual on the right.} \label{fig:figrandomAC}
\vspace{0.2in}
\end{figure}

Grangetto's Randomized Arithmetic Coding is one of the earliest
attempts to provide both source coding and encryption using
Arithmetic Coding~\cite{Grangetto04}. The idea of Randomized
Arithmetic Coding is to randomly swap (or not swap) the two
intervals corresponding to the symbols (`0' and `1') at every
iteration based on a random private binary key stream ($K$) which is
available only to the decoder of the intended party. This randomizes
the location of the final interval while retaining compression
efficiency. Having already established that Arithmetic Coding is a
specific mode of GLS, we can interpret Randomized Arithmetic Coding
as a swapping between two modes of the GLS (the two modes are duals
of each other) at every iteration based on a private key stream
(Fig.~\ref{fig:figrandomAC}).

Two important points that need to be remembered in randomized
arithmetic coding are:

\begin{enumerate}
\item The key stream $K$ should be perfectly random to ensure best
security.
\item The key stream $K$ is as long as the uncompressed message $M$.
\end{enumerate}

\begin{figure}[!h]
\centering
\includegraphics[scale=.6]{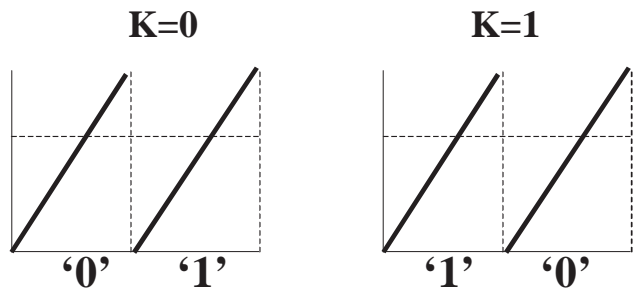}
\caption{OTP is a special case of Grangetto's Randomized Arithmetic
Coding performed on the binary map instead of the skewed binary map.
Thus OTP can be seen as finding the initial condition under random
switching on the binary map and its dual.} \label{fig:figOTP}
\vspace{0.2in}
\end{figure}

\subsection{OTP is a special case of randomized arithmetic coding}
It turns out that OTP encryption and decryption can be seen as
special case of randomized arithmetic coding. Instead of using the
skewed binary map, if we used the binary map then we end up with the
OTP encryption which Shannon showed in 1949 to be perfectly
secure.\\

\noindent {\bf Theorem:~} OTP encryption is equivalent to finding
the initial condition for the symbolic sequence $M$ under switching
based on key $K$ on the binary map and its dual.

\noindent {\bf Proof:~} We shall prove that this is equivalent to an
XOR operation between the message $M$ and the key $K$. Since XOR
operation is equivalent to OTP encryption, we would thus have a
proof of the theorem.

Let us consider all possibilities for one bit of the key $K$ and one
bit of the message $M$. When $K=0$ and $M=0$, the interval is mapped
to [0,0.5). The initial condition is going to lie in this interval
irrespective of future bits (this is because the map is continuous
in each of the intervals). The first bit of the initial condition is
going to be 0. When $K=0$ and $M=1$, the initial condition will lie
in [0.5,1) which would mean that the first bit of the initial
condition is 1. When $K=1$, the outputs are reversed. The output is
shown in Table~\ref{table:tab1}. It can be seen that this is
equivalent to the XOR operation between $K$ and $M$. Subsequent bits
would follow the same logic (one can imagine that the first bit of
the initial condition has been flushed as output and the interval
has been rescaled to [0,1) to begin encoding the second bit of $M$
with the second bit of $K$ for switching). $\hfill \square$

\begin{table}[!h]
\centering
 \caption{Switching between the two maps is equivalent to
XOR between $K$ and $M$.}\label{table:tab1} \vspace{0.1in}
\begin{tabular}{|c|c|c|}
  \hline
 $K$  & $M$ & First bit of initial condition \\
 \hline
  0 & 0 & 0 \\
  0 & 1 & 1 \\
  1 & 0 & 1 \\
  1 & 1 & 0 \\
  \hline
\end{tabular}
\end{table}

\begin{figure}[!h]
\centering
\includegraphics[scale=.6]{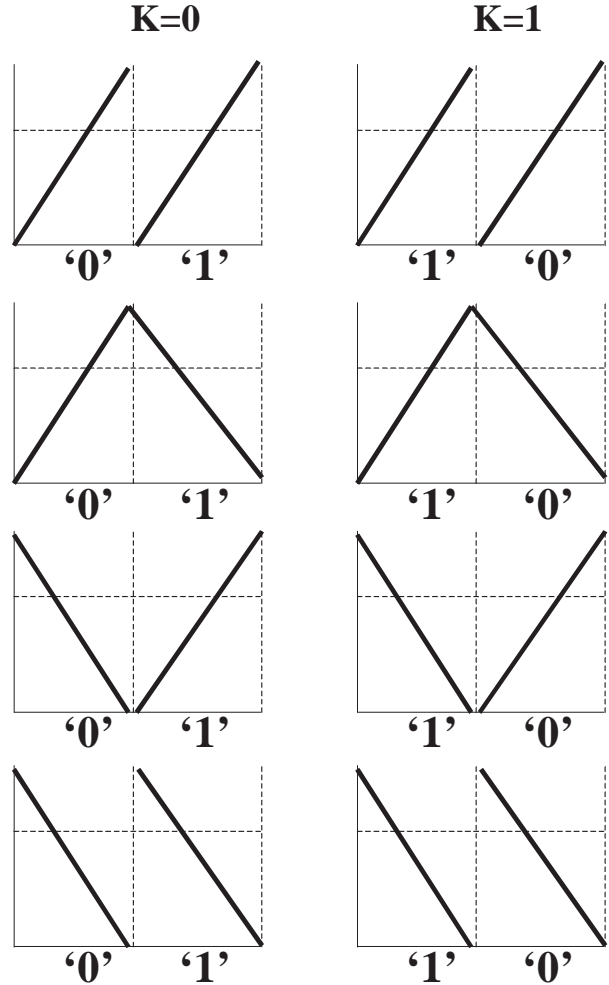}
\caption{Perfect secrecy systems equivalent to OTP. There is a
choice of 4 modes for $K=0$ and $K=1$ independently. Thus there are
16 possible perfect secrecy systems which are equivalent to OTP.}
\label{fig:figOTPmodes} \vspace{0.2in}
\end{figure}

We are effectively attempting to {\it compress} the message stream
$M$ under the switching operation. We are performing GLS-coding (or
arithmetic coding) on the binary map and its dual. However, it must
be noted that since the two intervals for the binary map and its
dual are of equal length, no compression will be achieved by the
method. Thus the initial condition when expressed in binary need to
have the same length as the message stream $M$ to enable {\it
lossless} decompression (in this case, decryption). Decryption
involves finding the symbolic sequence on the binary map (and its
dual under the operation of switching key $K$) using the initial
condition.

\subsection{Other perfect secrecy systems that are equivalent to OTP}
This connection between arithmetic coding, binary map and OTP
enables us to find perfect secrecy systems which are all equivalent
to OTP. We know that the binary map has 4 possible modes that
correspond to choosing either positive or negative slope in the two
intervals and 4 other modes that are duals. In order to obtain
secrecy systems that are equivalent to OTP, we can choose any of the
4 modes for $K=0$ and $K=1$ independently. Thus there are 16
possible secrecy systems which are all perfectly secure. The OTP is
in fact one of them (Fig.~\ref{fig:figOTPmodes}).

\section{$n$-$OTP$: generalization to non-binary alphabets}
\begin{figure}[!h]
\centering
\includegraphics[scale=.6]{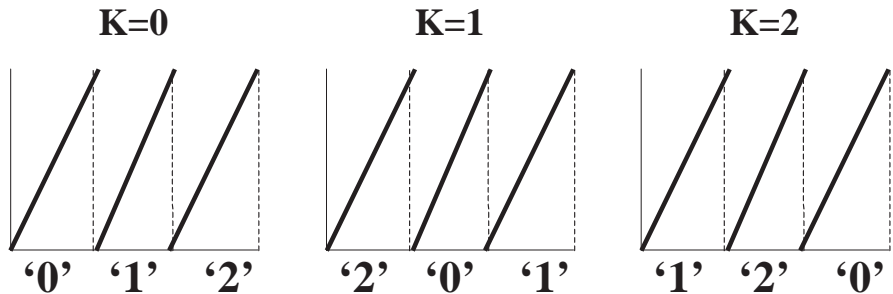}
\caption{$n$-$OTP$: For $n=3$ and for an input ternary message $M$,
the OTP can be implemented as finding the initial condition by
switching between the above three dynamical systems based on a
random key $K$ drawn from the alphabet \{0,1,2\} with equal
probabilities. This is equivalent to addition modulo 3 between $M$
and $K$.} \label{fig:fig3OTP} \vspace{0.2in}
\end{figure}

The dynamical system viewpoint that we have proposed immediately
enables us to generalize OTP for non-binary alphabets ($n$-$OTP$
where $n \geq 2$) while retaining perfect secrecy. Suppose we have a
message that takes values from the ternary alphabet $\{0,1,2\}$. We
further assume that we have a {\it perfect} random key stream that
also takes values from the ternary alphabet. To perform encryption,
we switch between the three GLS maps shown in Fig.~\ref{fig:fig3OTP}
depending on the key value.

For $n>2$, there are multiple options for choosing the $n$ dynamical
systems to switch. Thus it is possible to generalize OTP encryption
to larger alphabets.

\section{A Dynamical Systems Implementation of XOR and other Logic Gates}
As we noted earlier, it is well known that the OTP encryption is an
XOR (exclusive-OR) operation between message stream $M$ and the key
stream $K$. This means that we have a dynamical system
implementation of the XOR gate. Is it possible to get other logic
gates from this framework?

\subsection{XNOR, OR, AND, NAND, NOR and NOT gates}
\begin{figure}[!h]
\centering
\includegraphics[scale=.6]{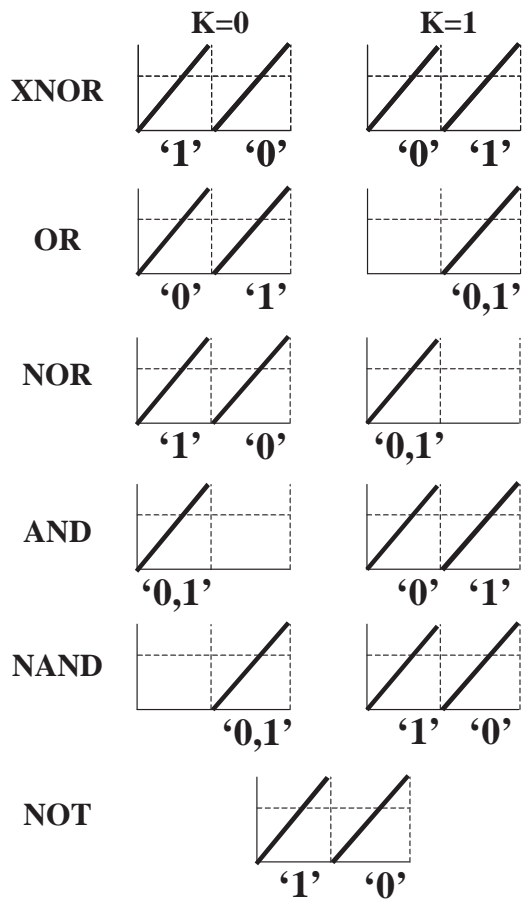}
\caption{XNOR, OR, NOR, AND and NAND logic gates implemented as
finding the initial condition by switching of dynamical systems. NOT
gate is just finding the initial condition on the single dynamical
system shown.} \label{fig:figlogicgates} \vspace{0.2in}
\end{figure}

We show that it is possible to implement the well known logic gates
XNOR, OR, AND, NAND, NOR and NOT gates as finding the initial
condition by switching of {\it appropriate} dynamical systems.
Please see Fig.~\ref{fig:figlogicgates} for a description of the
implementation of logic gates.

The NOT gate does not involve a switching operation. The NOT
operation can be seen as finding the initial condition on the binary
map with the symbols for the two intervals swapped. One can easily
extend this method for any logical gate. It is also possible to
extend these systems to higher alphabets just like how we did for
OTP. The ternary-OTP described in Section 5 would corresponds to
addition modulo 3 of $M$ and $K$.

There has already been substantial research on logic gate
implementation by a dynamical system approach and the building of a
chaotic computer (a computer where the components are
non-linear)~\cite{Sinha1, Sinha2, Sinha3}.  An advantage our
implementation has is that the input can be a stream of symbols
which can be buffered and the logical output which is an initial
condition (a real number who's binary representation is the logical
output) can also be stored and the output can be given as a stream
instead of performing the operation for every bit. In a sense, the
dynamical system can be used both to {\it perform} the logical
operation and {\it accumulate and store} the logical output
simultaneously. It remains to be seen whether the new
implementations proposed in this paper can be deployed in the
hardware in an efficient manner (fast, precise, compact and
low-power implementation).

\section{Conclusions}
We have established that {\bf finding the initial condition under
switching of dynamical systems belonging to the class of Generalized
Lur\"{o}th Series} acts as an unifying framework for lossless
compression, randomized arithmetic coding, perfect secrecy systems
and implementation of logic gates.

\end{document}